\documentclass[aps, prl, twocolumn, superscriptaddress, amsmath, showpacs]{revtex4-1}
\usepackage{graphicx}
\usepackage{csquotes}
\usepackage[usenames,dvipsnames]{color}
\usepackage{soul}
\usepackage{SIunits}

\newcommand{\dd}{\mathrm{d}}

\newcommand{\bet}{\emph}
\newcommand{\ord}{\mathcal{O}}
\newcommand{\dissip}{\epsilon}
\renewcommand{\vec}{\mathbf}

\newcommand{\picheight}{0.3\textheight}

\begin{document}

\title{Time-symmetry breaking in turbulence}

\author{Jennifer Jucha}\affiliation{Max Planck Institute for Dynamics and Self-Organization (MPIDS), 37077 G\"ottingen, Germany}
\author{Haitao Xu}\affiliation{Max Planck Institute for Dynamics and Self-Organization (MPIDS), 37077 G\"ottingen, Germany}
\author{Alain Pumir}\affiliation{Max Planck Institute for Dynamics and Self-Organization (MPIDS), 37077 G\"ottingen, Germany}
\affiliation{Ecole Normale Sup\'erieure de Lyon and CNRS, 69007 Lyon, France}
\author{Eberhard Bodenschatz}\affiliation{Max Planck Institute for Dynamics and Self-Organization (MPIDS), 37077 G\"ottingen, Germany}\affiliation{Institute for Nonlinear Dynamics, 37077 G\"ottingen, Germany}
\affiliation{Laboratory of Atomic and Solids State Physics and Sibley School for Mechanical and Aerospace Engineering,  Cornell University, Ithaca, NY, 14853 USA}

\begin{abstract}
In three-dimensional turbulent flows, the flux of energy from large to small scales breaks time symmetry. We show here that this irreversibility can be quantified  by following the relative motion of several Lagrangian tracers. 
We find by analytical calculation,  numerical analysis and experimental observation that the existence of the energy flux implies that, at short times, two particles separate temporally slower forwards than backwards, and the difference between forward and backward dispersion grows as $t^3$. We also find the geometric deformation of material volumes, surrogated by four points spanning an initially regular tetrahedron, to show sensitivity to the time-reversal with an effect growing linearly in $t$. We associate this with the structure of the strain rate in the flow. 
\end{abstract}

\pacs{47.27.Ak,47.27.Jv,47.27.Gs,47.27.tb}

\maketitle

In turbulent flows, far from boundaries, energy flows from the scale at which it is injected, $l_I$, to the scale where it is dissipated, $l_D$. For intense three-dimensional turbulence, $l_D \ll l_I$, and the energy flux, $\dissip$, is from large to small scales \cite{frisch95}. As a consequence, time symmetry is broken, since the  time reversal  $t \rightarrow -t$ would also reverse the direction of the energy flux. Exploring the implications of this time asymmetry on the relative motion between fluid particles is the aim of this Letter.\par

The simplest problem in this context concerns the dispersion of two particles whose positions, $\mathbf{r}_1(t)$ and $\mathbf{r}_2(t)$, are separated by $\mathbf{R}(t) =  \mathbf{r}_2(t) - \mathbf{r}_1(t)$. The growth of the mean squared separation, $\langle \mathbf{R}^2 (t) \rangle$, forwards ($t>0$) and backwards in time ($t<0$) is a fundamental question in turbulence research~\cite{R26} and is also related to important problems such as turbulent diffusion and mixing~\cite{S01,SC09}. At long times, both for $t > 0$ and $t < 0$, it is expected that the distance between particles increases according to the Richardson prediction as $ \langle \mathbf{R}^2 (t) \rangle \approx g_{f,b} \dissip |t|^3$~\cite{SC09}, with two constants, $g_f$ and $g_b$, for forward and backward dispersion, respectively. The lack of direct evidence for the Richardson $t^3$ regime in well-controlled laboratory flows~\cite{B06a} or in Direct Numerical Simulations (DNS)~\cite{SC09,BIC14} makes the determination of the constants $g_f$ and $g_b$ elusive, although it is expected that $g_b > g_f$~\cite{SC09,SYB05,B06}.\par

In this Letter we show that \bet{for short times} the flow irreversibility imposes a quantitative relation between forward and backward particle dispersion. For particle pairs, the energy flux through scales is captured by
\begin{equation}
\left\langle \frac{d}{dt} \left[\mathbf{v}_2(t) - \mathbf{v}_1(t) \right]^2 \Big|_0 \right\rangle = - 4 \dissip,
\label{eq:flux_lag}
\end{equation}
where $\mathbf{v}_{1}(t) $ and $\mathbf{v}_{2}(t)$  are the Lagrangian velocities of the particles and the average is taken over all particle pairs with the same initial separation, $ | \mathbf{R}(0) | =R_0$, in the inertial subrange ($l_D \ll R_0 \ll l_I$). Equation (\ref{eq:flux_lag}) is exact in the limit of very large Reynolds number~\cite{MOA99,FGV01,PSC01} and can be seen as the Lagrangian version of the Kolmogorov 4/5-law~\cite{frisch95}. For short times, Eq.~(\ref{eq:flux_lag}) implies that backward particle dispersion is faster than the forward case, with
\begin{equation} 
\langle \mathbf{R}^2(-t) \rangle - \langle \mathbf{R}^2(t) \rangle = 4 \dissip t^3 + \ord(t^5).
\label{eq:diff_bac_for}
\end{equation}
The $t^3$ power in Eq.~(\ref{eq:diff_bac_for}) is strongly reminiscent of the Richardson prediction, with the expectation that $g_b > g_f$ at longer times. The relation between the irreversibility predicted by Eq.~(\ref{eq:diff_bac_for}) and the one expected at longer times ($g_b > g_f$), however, remains to be established.\par

Whereas the difference between backward and forward pair dispersion at short times is weak ($\propto t^3$), we found a strong  manifestation of the time asymmetry when investigating multi-particle dispersion. 
The analysis of the deformation of an initially regular tetrahedron consisting of four tracer particles~\cite{PSC00,XOB08} reveals a stronger flattening of the shape forwards in time, but a stronger elongation backwards in time. 
We relate the observed time asymmetry in the shape deformation to a fundamental property of the flow~\cite{Betchov56,Siggia81,Ashurst87,Pumir13} by investigating the structure of the perceived rate of strain tensor based on the velocities of the four Lagrangian particles~\cite{XPB11}.\par

Our finding relies on analytical calculation, DNS, and data from  3D Lagrangian particle tracking in a laboratory flow. 
The experiments were conducted with a von K\'arm\'an swirling water flow.  The setup consisted of a cylindrical tank with a diameter of $\unit{48.3}{\centi\meter}$  and a height of $\unit{60.5}{\centi\meter}$, with counterrotating impellers installed at the top and bottom. Its geometry is very similar to the one described in Ref.~\cite{O06}, but with a slightly different design of the impellers to weaken the global structure of the flow. 
At the center of the tank, where the measurements were performed, the flow is nearly homogeneous and isotropic. 
As tracers for the fluid motion, we used polystyrene microspheres of density $\rho=1.06\, \rho_{\text{water}} $ and a diameter close to the Kolmogorov length scale, $\eta$. 
We measured the trajectories of these tracers using Lagrangian particle tracking with sampling rates exceeding 20 frames per Kolmogorov time scale, $\tau_\eta$~\cite{O06a,X08}. 
We obtained three data sets at $R_\lambda=270$, $350$ and $690$, with corresponding Kolmogorov scales $\eta=\unit{105}{\micro\meter}$, $\unit{66}{\micro\meter}$, and $\unit{30}{\micro\meter}$ and $\tau_\eta=\unit{11.1}{\milli\second}$, $\unit{4.3}{\milli\second}$, and $\unit{0.90}{\milli\second}$, respectively. 
The integral length scales of $L\approx\unit{5.5}{\centi\meter}$ for the first two and $L\approx\unit{7.0}{\centi\meter}$ for the last data set are both smaller than the size of the measurement volume, which is approximately $(\unit{8}{\centi\meter})^3$.
Many independent, one-second recordings of $\sim 100$ particles where combined to generate sufficient statistics.
For example, the $R_\lambda = 690$ dataset contains 555,479 particle trajectories lasting at least $20 \tau_\eta$.
Our experimental results are compared to DNS data obtained from pseudo-spectral codes~\cite{vosskuhle:2013,li2008,Y12}.

To study the dispersion between two particles, it is more convenient to analyze the change in separation, $\delta \mathbf{R}(t) = \mathbf{R}(t) - \mathbf{R}(0)$, than the separation $\mathbf{R}(t)$ itself~\cite{B50,O06,SC09}. We expand $\delta \mathbf{R}(t)$ in a Taylor series and average over many particle pairs with a fixed initial separation $ | \mathbf{R}(0)|=R_0$ to obtain
\begin{equation}
\frac{\langle \delta \mathbf{R}(t)^2\rangle}{R_0^2} =  
\frac{\langle \mathbf{u}(0)^2\rangle}{R_0^2} t^2 
+ \frac{\left\langle \mathbf{u}(0) \cdot \mathbf{a}(0) \right\rangle}{R_0^2}  t^3 
+ \ord(t^4) ,
\label{eq:evol_dR2}
\end{equation}
where $\mathbf{u}(0)$ and $\mathbf{a}(0)$ are the relative velocity and acceleration between the two particles at time $t=0$. Using Eq.~\eqref{eq:flux_lag} reduces the $t^3$ term in Eq.~\eqref{eq:evol_dR2} to $-2 (t/t_0)^3$, where $t_0 = (R_0^2/\dissip)^{1/3}$ is the (Kolmogorov) time scale characteristic of the motion of eddies of size $R_0$~\cite{frisch95}. Eq.~\eqref{eq:evol_dR2} can thus be expressed as
\begin{equation}
\frac{\langle \delta \mathbf{R}(t)^2\rangle}{R_0^2} =   \frac{\langle \mathbf{u}(0)^2\rangle}{(\dissip R_0)^{2/3}} \Bigl( \frac{t}{t_0} \Bigr)^2 - 2 \Bigl(\frac{t}{t_0} \Bigr)^3 + \ord(t^4).
\label{eq:evol_dR2_nodim}
\end{equation}
For short times, the dominant behavior is given by the $t^2$ term in Eq.~\eqref{eq:evol_dR2_nodim}~\cite{B50}, which is even in $t$, and thus reveals no asymmetry in time. The odd $t^3$ term is the first to break the $ t \rightarrow -t$ symmetry. This is better seen from the difference between the forward and backward dispersion,
\begin{align}
\frac{\langle \delta \mathbf{R}(-t)^2- \delta \mathbf{R}(t)^2\rangle}{R_0^2} 
& = -2 \frac{\left\langle \mathbf{u}(0) \cdot \mathbf{a}(0) \right\rangle}{R_0^2} t^3 + \ord (t^5) \nonumber \\
&= 4 ({t}/{t_0})^3 + \ord (t^5),
\label{eq:Rb_Rf}
\end{align}
which is equivalent to Eq.~\eqref{eq:diff_bac_for}. We note that the simple form of Eq.~\eqref{eq:evol_dR2_nodim}, which suggests that the evolution of $\langle \delta \mathbf{R}^2(t) \rangle$ depends on $(t/t_0)$ alone, is accurate only up to $\ord(t/t_0)^3$. Not all higher-order terms in the Taylor expansion can be reduced to functions of $(t/t_0)$~\cite{F13}.\par

To test Eq.~\eqref{eq:Rb_Rf}, we identified particle pairs from our large set of experimental and numerical trajectories with a given initial separation $R_0$  and studied the evolution of $\delta \mathbf{R}(t)^2$, both forwards and backwards in time.  One of the difficulties of reliably measuring $\langle \delta \mathbf{R}(t)^2 \rangle$ in experiments comes from the finite size of the measurement volume in which particles are tracked. The residence time of particle pairs in the measurement volume decreases with the separation velocity, inducing a bias. We analyze how this affects the results in the Appendix and show that the effect is weak. 
The very good agreement between experiments and DNS convinces us  that the finite-volume bias does not alter our results.\par

\begin{figure}[t]
\includegraphics[height=\picheight]{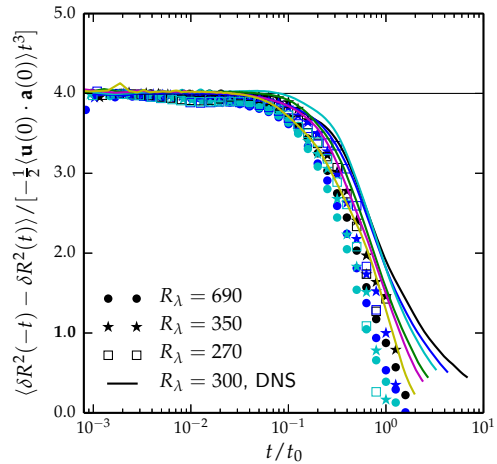}
\caption{(color online). The difference between the backward and forward mean squared relative separation, $\langle \delta \mathbf{R}(-t)^2 - \delta \mathbf{R}(t)^2 \rangle$, compensated using Eq.~\eqref{eq:Rb_Rf}. 
The symbols correspond to experiments:
circles for $R_\lambda = 690$ ($R_0/\eta = 267,\,333,\,400$), 
stars for $R_\lambda = 350$ ($R_0/\eta = 152,\,182,\,212$), 
and squares for $R_\lambda = 270$ ($R_0/\eta = 95,\,114,\,133$). 
The lines correspond to DNS at $R_\lambda=300$ ($R_0/\eta=19,\,38,\,58,\,77,\,92,\,123$).}
\label{figThirdOrder}
\end{figure}

Fig.~\ref{figThirdOrder} shows the difference, $\langle \delta \mathbf{R}^2(-t) -  \delta \mathbf{R}^2(t) \rangle$, compensated by 
$- \frac{\left\langle \mathbf{u}(0) \cdot \mathbf{a}(0) \right\rangle}{2 R_0^2} t^3 $, 
using Eq.~\eqref{eq:Rb_Rf}, obtained from both experiments and DNS at 4 different Reynolds numbers. 
The DNS, $R_\lambda = 300$ data consisted of $32,768$ particle trajectories in
a statistically stationary turbulent flow~\cite{vosskuhle:2013} over $\sim 4.5$ large-eddy turnover times, allowing 
particle pairs with a prescribed size to be followed for a long period of time.
The data all show a clear plateau up to $t\approx t_0/10$, in complete agreement with Eq.~\eqref{eq:Rb_Rf}. 
At longer times, both experimental and DNS data decrease rapidly towards zero without any sign of the plateau expected from the Richardson prediction,
\begin{equation}
\frac{\langle \delta \mathbf{R}(-t)^2- \delta \mathbf{R}(t)^2\rangle }{R_0^2} =(g_b - g_f) \Bigl(\frac{t}{t_0} \Bigr)^3 .
\label{eqRichardson}
\end{equation}
While the slightly faster decay of the experimental data for $t \gtrsim t_0$ could be due to a residual finite-volume bias, this should not affect the DNS data. Previous experiments at $R_\lambda = 172$ with initial separations in the range $4 \le R_0/\eta \le 28$ suggested a value of the difference of $(g_b - g_f) = 0.6 \pm 0.1$~\cite{B06}. 
Fig.~\ref{figThirdOrder} does not provide evidence for this value, although it does not rule out the existence of a plateau at a lower value of  $(g_b - g_f)$.  Note that Eq.~\eqref{eqRichardson} predicts the time irreversibility caused by the energy flux to persist into the inertial range and remarkably to grow as $t^3$ as well. It is therefore tempting to draw an analogy between Eq.~\eqref{eq:Rb_Rf}, which is exact and valid at short times, and the expected Richardson regime at longer times~\cite{B12}. The fact that a plateau corresponding to $(g_b - g_f)$ would be substantially lower than the value of $4$ given by Eq.~\eqref{eq:Rb_Rf} indicates that the connection between the short-time behavior, Eq.~\eqref{eq:Rb_Rf}, and the longer-time behavior, Eq.~\eqref{eqRichardson},
requires a deeper understanding.\par

The time irreversibility predicted by Eq.~\eqref{eq:Rb_Rf} for particle pair separations grows slowly at small times, $\propto t^3$. We discuss below a stronger ($\propto t$) manifestation of the time irreversibility by analyzing the evolution of four particles initially forming a regular tetrahedron. Additionally, the motion of tetrahedra provides insight into the structure of a flow~\cite{CPS99,PSC00,XOB08,XPB11,Pumir13} and in fact into the origin of the irreversibility observed in particle pair separation.\par

The geometry of a set of four points $(\vec{x}_1, ... \vec{x}_4)$, i.e., a tetrahedron, can be effectively described by three vectors. The position of the tetrahedron is immaterial in a homogeneous flow.
The shape tensor, $G_{ij} = \sum_a (x_{a,i} - x_{C,i})(x_{a,j} - x_{C,j})$, where $x_{a,i}$ is the $i^{th}$ component of $\vec{x}_a$, provides an effective description of the tetrahedron geometry. The radius of gyration of the tetrahedron, $R^2(t) = \text{tr}(\mathbf{G})=\frac14 \sum_{a<b} |\vec{x}_a(t) - \vec{x}_b(t)|^2$, is simply given by the trace of $\mathbf{G}$. The shape is described by the three eigenvalues $g_i$ of $G$, with $g_1\geq g_2 \geq g_3$. For a regular tetrahedron, where all edges have the same length, all three eigenvalues are equal.  For $g_1 \gg g_2\approx g_3$, the tetrahedron is needle-like,  while $g_1\approx g_2 \gg g_3$ represents a pancake-like shape.\par

\begin{figure}[t]
\includegraphics[height=\picheight]{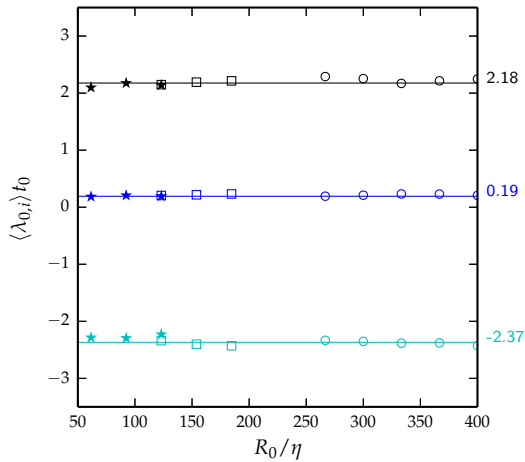}
\caption{(color online). Eigenvalues of the perceived rate-of-strain tensor, 
$\lambda_{0,i} t_0$, $(i=1,\,2,\,3)$, defined on tetrahedra with different sizes $R_0 /\eta$.
Open symbols are from experiments at $R_\lambda= 690$ and $350$ and filled symbols from DNS at $R_\lambda=300$.  
The solid lines are the corresponding averages for $i=1$ (top), $2$ (middle), and $3$ (bottom).}
\label{figStrain}
\end{figure}

The evolution of $\mathbf{G}$ can be conveniently written in the compact form~\cite{Pumir13}
\begin{equation}
\frac{\dd }{\dd t}\mathbf{G}(t) = \mathbf{M}(t) \mathbf{G}(t) + \mathbf{G}(t) \mathbf{M}^T(t) ,
\label{eq:dG_dt}
\end{equation}
where $\mathbf{M}(t)$ is the perceived velocity gradient tensor that describes the turbulent flow field seen by the 4 points~\cite{CPS99,XPB11}. The perceived velocity gradient reduces to the usual velocity gradient when the tetrahedron becomes smaller than the Kolmogorov scale, $\eta$~\cite{Pumir13}. We solve Eq.~\eqref{eq:dG_dt} for short times using a Taylor expansion around $t=0$ and taking $G_{ij}(0) = (R_0^2/2) \delta_{ij}$ as the initial condition, i.e., the tetrahedra are initially regular with edge lengths $R_0$. The solutions for the average size and shape are
\begin{align}
\langle R^2(t) \rangle & = \frac{R_0^2}{2} \bigg[3 + 2 \text{tr} \langle \mathbf{S}_0^2\rangle t^2 \nonumber\\
& \quad + 2 \text{tr}\left( \frac23 \langle \mathbf{S}_0^3 \rangle +\langle\mathbf{S}_0 \mathbf{\dot{S}}_0 \rangle \right) t^3 + \ord(t^4) \bigg]
\label{eqRadius}
\end{align}
and 
\begin{align}
\langle g_i \rangle &=  \frac{R_0^2}{2} \bigg[1 + 2 \langle \lambda_{0,i} \rangle t \nonumber\\
& \quad + \left( 2 \langle \lambda_{0,i}^2 \rangle + \langle \mathbf{\dot{S}}_{0,ii} \rangle \right) t^2 + \ord(t^3) \bigg]. 
\label{eqEigen}
\end{align}
At the orders considered, the evolution of the tetrahedron geometry depends only on the perceived rate-of-strain tensor, $\mathbf{S}_0 = \mathbf{S}(0) = \frac12 [ \mathbf{M}(0) +\mathbf{M}(0)^T]$, whose eigenvalues, $\lambda_{0,i}$, are sorted in decreasing order ($\lambda_{0,1} \ge \lambda_{0,2} \ge \lambda_{0,3}$), and on its time-derivative, $\mathbf{\dot{S}}_0 = \frac{\dd}{\dd t} \mathbf{S}(t)\big|_0$. In Eq.~\eqref{eqEigen}, all terms are in fact expressed in the eigenbasis of $\mathbf{S}_0$. \par 

\begin{figure*}[t]
\centering
\includegraphics[height=\picheight]{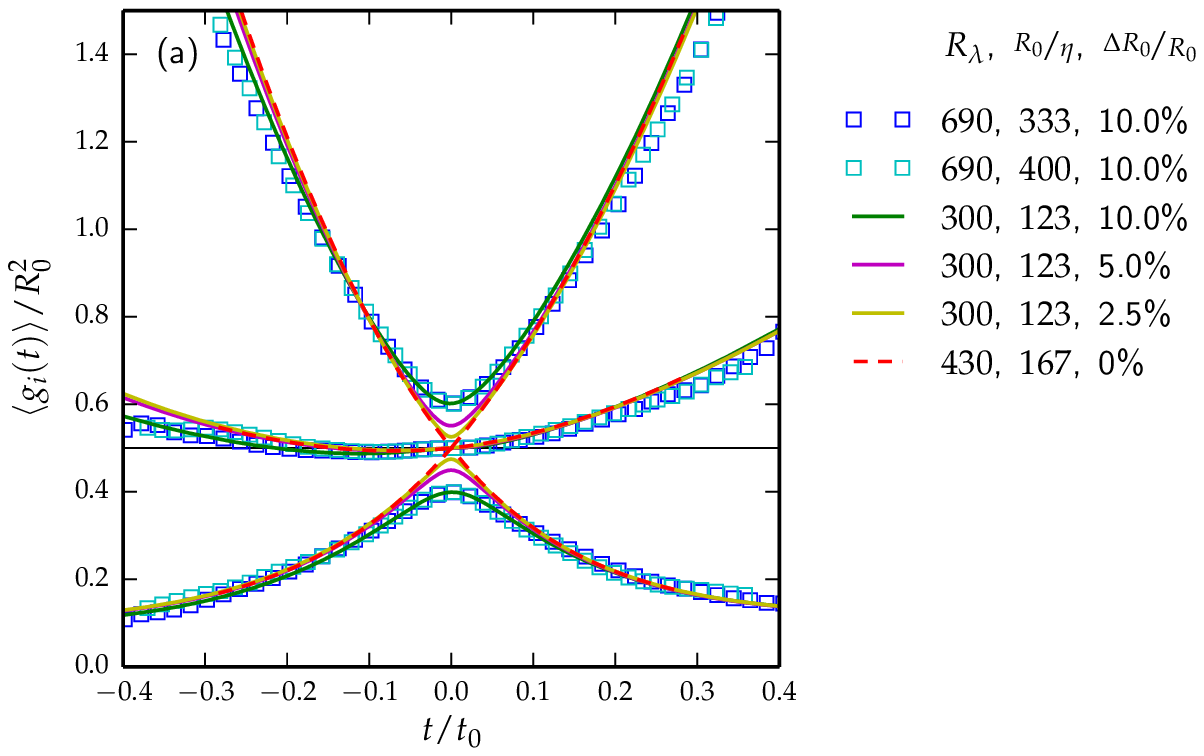}
\includegraphics[height=\picheight]{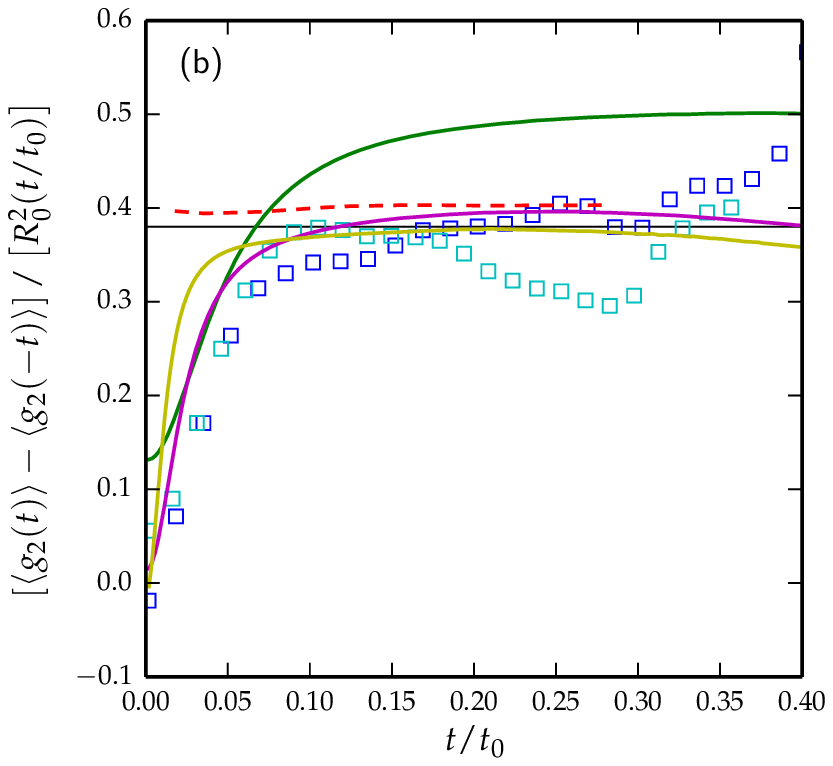}
\caption{(color online). Time-asymmetry revealed from the geometry of tetrahedra. (a) The eigenvalues of the shape tensor $\mathbf{G}$, $\langle g_i (t) \rangle$, divided by $R_0^2$, such that initially regular tetrads with size $R_0$ start at $g_i(0)/R_0^2 = 1/2$. (b) The difference in the backward and forward growth rates of the intermediate eigenvalue, $[ \langle g_{2}(t) - g_{2}(-t) \rangle] /[R_0^2 (t/t_0)]$, behaves as $2 \langle  \lambda_{2,0} \rangle t_0$ to leading order according to Eq.~\eqref{figShape}. 
The horizontal thin black line indicates the value of $2 \langle \lambda_{0,2} \rangle t_0 \approx 0.38$ as given by Fig.~\ref{figStrain}. The experimental results ($R_\lambda = 690$) and the DNS at $R_\lambda = 300$, were obtained with tetrads close to regular, with a small tolerance $\Delta R_0$ on the edge lengths, indicated in the legend. The dashed lines (DNS, $R_\lambda = 430$) were obtained with regular tetrads. The effect of $\Delta R_0 \ne 0$ is the strongest at small $t$ and diminishes when $\Delta R_0/R_0$ decreases.}
\label{figShape}
\end{figure*}

We first note that the radius of gyration, $R^2(t)$, can also be expressed as an average over the squares of the edge lengths of the tetrahedron. Thus, Eq.~\eqref{eqRadius} must be consistent with Eq.~\eqref{eq:evol_dR2}.  This implies that $\text{tr} \langle \mathbf{S}_0^2\rangle  = \frac{3}{2 R_0^2} \langle \mathbf{u}(0)^2\rangle$  and  $\text{tr}\big( \frac23 \langle \mathbf{S}_0^3 \rangle _t+\langle\mathbf{S}_0 \mathbf{\dot{S}}_0 \rangle \big) = \frac32 \left\langle \mathbf{u}(0)\cdot\mathbf{a}(0) \right\rangle$, which we explicitly confirmed with our data. Furthermore, the incompressibility of the flow imposes that $\mathbf{M}$ (and hence $\mathbf{S}$) is traceless \bet{on average}, which means that $\langle \lambda_{0,1} \rangle \geq 0$ and $\langle \lambda_{0,3} \rangle \leq 0$. 
The generation of small scales by turbulent flows, which plays a key 
role in the energy cascade, implies that the intermediate eigenvalue of the rate of strain tensor is positive~\cite{Betchov56}.
This property also applies to the \bet{perceived} velocity gradient tensor in the inertial range~\cite{Pumir13} (Fig.~\ref{figStrain}). Remarkably, our data suggest that $\langle \lambda_{0,i} \rangle t_0 \approx \text{const}$ over the range of Reynolds numbers and inertial scales covered here. For initially regular tetrahedra of edge length $R_0$, Eq.~\eqref{eqEigen} predicts that $\langle g_i (t) \rangle = \frac12 R_0^2$ at $t=0$ and that $\langle g_i (t) \rangle$ grows \bet{linearly} as $R_0^2 \langle \lambda_{0,i} \rangle t$ for small $t$. 
The tetrahedra obtained experimentally and numerically at $R_\lambda=300$, however, are not strictly regular, but correspond to a set of 4 points whose relative distances are equal to within a fixed relative tolerance in the range $2.5 - 10 \%$.
Fig.~\ref{figShape}(a) shows that the linear behavior predicted by Eq.~\eqref{eqEigen} is observed when the tetrahedra are regular, as obtained using the Johns Hopkins University database~\cite{li2008,Y12} ($R_\lambda = 430$), or when the tolerance is reduced.
The time asymmetry in this shape evolution, seen from the eigenvalues of $\mathbf{G}$ in Fig.~\ref{figShape}, originates from the positive value of $\langle \lambda_{0,2} \rangle $.
For regular tetrahedra, Eq.\eqref{eqEigen} shows that in the eigenbasis of $\mathbf{S}_0$, the largest eigenvalue of $\mathbf{G}$ is $g_1$ for $t > 0$, and $g_3$ for $t < 0$. The difference between the largest eigenvalues at $t > 0$ (forwards in time) and at $t<0$ (backwards in time) is thus $R_0^2 \langle (\lambda_{0,1} + \lambda_{0,3}) t \rangle  = - R_0^2 \langle \lambda_{0,2} t \rangle$. 
In fact, the difference between the backward and forward growth rates of the intermediate eigenvalue, $\langle g_2 \rangle$, shows an even stronger asymmetry:
\begin{equation}
 \langle g_{2}(t) - g_{2}(-t) \rangle /[R_0^2 (t/t_0)] = 2 \langle  \lambda_{0,2} \rangle t_0 + \ord(t^2). 
\label{eq:g2diff}
\end{equation}
The expected plateau of $2 \langle \lambda_{0,2} \rangle t_0$ is 
seen in Fig.~\ref{figShape}(b) when the tetrads are regular, or when the tolerance on the initial edge lengths is reduced.\par

In summary, we have shown that the relative motion between several Lagrangian particles reveals  the fundamental irreversibility of turbulent flows. At short times, the time asymmetry of two-particle dispersion grows as $t^3$, which is deduced from an identity derived from the Navier-Stokes equations in the large $R_\lambda$ limit that expresses the existence of a downscale energy cascade. Our study, however, leaves open the question of the existence of two different constants governing the dispersion forwards and backwards in time in the Richardson regime~\cite{SYB05,B06}.
A stronger manifestation of the time asymmetry, $\propto t$, was observed by studying the shape deformation of sets of four points.
This asymmetry can be understood from another fundamental property of turbulence, namely the existence of a positive intermediate eigenvalue of the rate-of-strain tensor~\cite{Betchov56,Pumir13}. Thus, remarkably, the manifestations of irreversibility are related to fundamental properties of the turbulent flow field. 

The time-symmetry breaking revealed by multi-particle statistics is a direct consequence of the energy flux through spatial scales (see
also \cite{FF13}).
The very recently observed manifestation of irreversibility~\cite{XPFB14} when following only a single fluid particle, where an intrinsic length scale is lacking, thus presents an interesting challenge to extend the analysis presented here. 
We expect that further insights into the physics of turbulence can be gained by analyzing the motion of tracer particles.

\acknowledgments{
We acknowledge the support from the Max Planck Society. AP also acknowledges ANR (contract TEC 2), the Humboldt foundation, and the PSMN at the Ecole Normale Sup\'erieure de Lyon.}


%

\section{appendix} 
\begin{figure}[htb]
\includegraphics[width=0.46\textwidth]{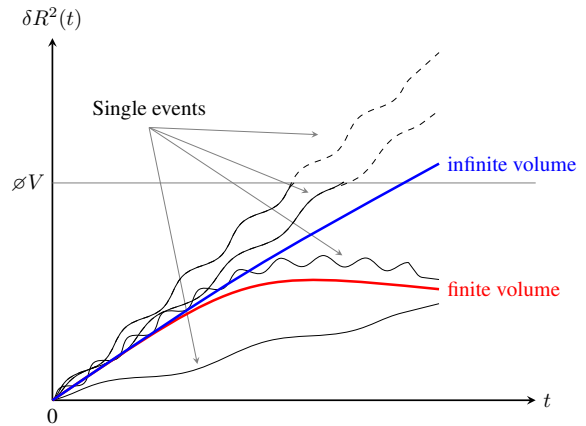}
\caption{(color online).  The blue curve shows the ensemble average for an infinite volume, the red curve the average over a time dependent ensemble for a finite volume. Black curves show examples of single events from these ensembles, with the dashed part not accessible in the case of a finite measurement volume.}
\label{figBiasSketch}
\end{figure}

The measurement volume in our experiment is finite and particles are thus only tracked for a finite time. The larger the relative velocity between two particles, $| \mathbf{u}(0)|$, the shorter they reside in the measurement volume~\cite{B06,LBOM07}. 
The experimentally measured mean squared displacement, $\langle \delta \mathbf{R}^2(t) \rangle_m$, determined by particle pairs which could be tracked up to time $t$, is smaller than the true value $\langle \delta \mathbf{R}^2(t) \rangle$ (see  Fig.~\ref{figBiasSketch}). \par

To quantitatively analyze this effect, we parametrize the bias in $\langle \delta \mathbf{R}^2(t) \rangle_m$ due to the loss of particles with large relative motions by generalizing Eq.~\eqref{eq:evol_dR2} to 
\begin{equation}
\frac{\langle \delta \mathbf{R}(t)^2\rangle_m}{R_0^2} =  
 \frac{\langle \mathbf{u}(0)^2\rangle}{(\dissip R_0)^{2/3}}  f_1(t) \Bigl( \frac{t}{t_0} \Bigr)^2 -
2 f_2(t)  \Bigl(\frac{t}{t_0} \Bigr)^3 + \ord(t^4).
\label{eqBias1}
\end{equation}
In Eq.~\eqref{eqBias1}, the functions $f_1(t)$ and $f_2(t)$ express that the values of the relative velocities of particles staying in the measurement volume for a time $t$ is {\it smaller} than the relative velocity of all particle pairs (see Fig.~\ref{figBiasSketch}). From our experimental data, we find that $f_i(t)>0.9$ for $t/t_0 <0.2$. Additionally, we restrict ourselves to particle pairs that can be tracked in the interval $[-t , t]$, ensuring that $f_i(t) = f_i(-t)$. We thus find that the time asymmetry between backward and forward dispersion is
\begin{equation}
\frac{\langle \delta \mathbf{R}(-t)^2- \delta \mathbf{R}(t)^2\rangle_m}{R_0^2} =4 f_2(t) \Bigl( \frac{t}{t_0} \Bigr)^3 +\ord(t^5). 
\label{eqBias2}
\end{equation}
The bias in Eq.~\eqref{eqBias2} is due only to the $f_2(t)$ term, and not to the leading term in Eq.~\eqref{eqBias1}. 
Over the short time interval where Fig.~\ref{figThirdOrder} shows a plateau, the error due to $f_2(t)$ is smaller than $\sim 10\%$.

\end{document}